\documentclass[12pt,final,english]{iopart}
\usepackage[dvips]{graphics}
\usepackage{floatflt}
\usepackage[T1]{fontenc}
\usepackage[latin1]{inputenc}
\usepackage{graphicx}
\usepackage{amssymb}
\usepackage{setspace}
\usepackage{color}
\usepackage{float}
\usepackage{babel}
\usepackage{dcolumn}
\usepackage{bm}
\usepackage{babel}

\newcommand {\ie}           {\textit{i.e.} }

\renewcommand {\etal}         {\textit{et al} }

\newcommand {\bv}           {\mathbf{v}}
\newcommand {\bE}           {\mathbf{E}}
\newcommand {\bB}           {\mathbf{B}}
\newcommand {\bA}           {\mathbf{A}}
\newcommand {\ba}           {\mathbf{a}}
\newcommand {\bb}           {\mathbf{b}}
\newcommand {\bn}           {\mathbf{n}}
\newcommand {\bh}           {\mathbf{h}}

\newcommand {\bdl}          {\mathbf{dl}}
\newcommand {\bX}           {\mathbf{X}}
\newcommand {\bJ}           {\mathbf{J}}
\newcommand {\curl}         {\nabla \times}
\newcommand {\grad}         {\nabla}
\newcommand {\vdiv}         {\nabla \cdot}
\newcommand {\F}            {\widetilde}
\newcommand {\bxi}          {\mbox{\boldmath{$\xi$}}}
\newcommand {\bldeta}       {\mbox{\boldmath{$\eta$}}}

\newcommand {\iotabar}      {\mbox{$\,\iota\!\!$-}}

\makeatother
\begin{document}

\title[Relaxed MHD states of a multiple region plasma]{Relaxed MHD states of a multiple region plasma}
  
\author{M. J. Hole\footnote[3]{To whom correspondence should be addressed
(matthew.hole@anu.edu.au)}, R. Mills, S. R. Hudson\dag and R. L. Dewar}
  
\address{Research School of Physics and Engineering, Australian National University, ACT 0200, Australia}
\address{\dag\  Princeton Plasma Physics Laboratory, P.O. Box 451, Princeton, New Jersey 08543, U.S.A.} 


\begin{abstract} 
We calculate the stability of a multiple relaxation region MHD (MRXMHD) plasma, or stepped-Beltrami plasma, using both variational and tearing mode treatments. 
The configuration studied is a periodic cylinder. 
In the variational treatment, the problem reduces to an eigenvalue problem for the interface displacements. 
For the tearing mode treatment, analytic expressions for the tearing mode stability parameter $\Delta'$, being the jump in the logarithm in the helical 
flux across the resonant surface, are found. 
The stability of these treatments is compared for $m=1$ displacements of an illustrative RFP-like configuration, 
comprising two distinct plasma regions. For pressure-less configurations, we find the marginal stability conclusions of each treatment
to be identical, confirming analytic results in the literature. 
The tearing mode treatment also resolves ideal MHD unstable solutions for which $\Delta' \rightarrow \infty$: these correspond to displacement of a resonant interface.
Wall stabilisation scans resolve the internal and external ideal kink. 
Scans with increasing pressure are also performed: these indicate that both variational and tearing mode treatments have the same stability trends with $\beta$,
and show pressure stabilisation in configurations with increasing edge pressure. 
Combined, our results suggest that MRXMHD configurations which are stable to ideal perturbations plus tearing modes are automatically in a 
stable state. Such configurations, and their stability properties, are of emerging importance in the quest to find
mathematically rigorous solutions of ideal MHD force balance in 3D geometry. 

\end{abstract}


\submitted
\noindent This is an author-created, un-copyedited version of an article submitted for publication in Nuclear Fusion on 12/01/09.

\maketitle
 

\section{Introduction}\label{sec:intro} 

Recently, Hole \etal \cite{Hole_07} proposed a model for a partially relaxed plasma-vacuum system. The purpose of the model,
which abandons all but a small number of flux surfaces, is to 
provide a mathematically rigorous foundation for ideal MHD equilibria in 3D configurations.
The model appeals to both chaotic field lines, that flatten the pressure gradient in chaotic regions, and 
Taylor relaxation, which force the plasma gradient to be zero in Taylor-relaxed regions. The model consists
of a stepped pressure profile, where the steps correspond to ideal MHD barriers across which can be 
supported a pressure or field jump, or a jump in rotational transform. Our overarching objective is the development of an
equilibrium solver for 3D plasmas built on a stepped pressure profile model. In the 3D case, we envisage that
the barriers can be chosen to be non-resonant KAM surfaces that survive the onset of field line chaos intrinsic to
3D equilibria. In between the interfaces, the field is Beltrami, such that $\curl \bB = \mu \bB$. The boundary condition
across the interfaces is the continuity of total pressure $p + B^2/2\mu_0$. 
Such a model, which we term a MRXMHD (multiple relaxation regions MHD) model, 
raises a number of questions. How should the equilibrium be constrained? How much jump in pressure and/or rotational transform $\iotabar$ 
can each interface support? Are the interfaces stable to deformation? Can the class of stability shed information
onto other quasi-relaxed phenomena? 

Previous work has focused on the equilibrium constraints \cite{Hole_06, Kukushkin_99}, construction of a 
numerical algorithm for calculation of Beltrami fields between interfaces in 3D configurations \cite{Hudson_07},
and a variational principle for the equilibrium and stability of the multiple interface configuration in cylindrical plasmas \cite{Hole_07}. 
We have also explored the relationship between relaxed plasma equilibrium models discussed here, and entropy related
plasma self-organisation principles \cite{Dewar_08}.
The motivation of this paper is to understand the nature of MRXMHD modes identified from the variational principle. 
Our work complements a separate in-press publication \cite{Mills_08} that unifies relaxed and ideal MHD principles
for constructing global solutions comprising mixed relaxed and ideal regions. 

Recently, Tassi \etal \cite{Tassi_07}, performed a tearing mode stability treatment on stepped $\mu$ force-free equilibria close to Taylor relaxed states. 
Their motivation was to develop a mechanism for the formation of cyclic Quasi-Single-Helicity (QSH) states
observed in Reverse Field Pinches \cite{Martin_07}.
They consider a cylindrical plasma divided into two different Beltrami regions, and encased in a perfectly conducting shell,
and compute the tearing mode stability parameter
$\Delta'$ at a resonant radius $r_s$ for a helical flux perturbation $\chi_1(r) = m B_{z1}(r) - \kappa r B_{\theta 1}(r)$.
Here, $r$ is the radial coordinate, and $m$ and $\kappa$ the poloidal and axial wave number.
Tassi \etal find critical values of the jump in $\mu$, beyond which the RFP-like plasma is unstable.
Based on these, they postulate the QSH state may be viewed as a small, cyclic departure from a Taylor-relaxed state. 

In this work, we extend the tearing mode stability treatment of Tassi \etal \cite{Tassi_07} to plasmas with finite pressure and a vacuum region, 
and compare stability conclusions of our variational treatment to that of a tearing mode stability analysis. 
Our paper is arranged as follows : Sec. 2 summarises the variational model of stepped pressure profile plasmas, presented in Hole \etal \cite{Hole_07}, 
and introduces a tearing mode model.  
Section 3 treats MRXMHD plasmas in cylindrical geometry, yielding stability parameter expressions 
for both the variational and tearing mode treatments. 
In Sec. 4, we compute stability for an example configuration, draw comparisons between the stability conclusions
based on variational and tearing mode treatments, and explore marginal stability limits in wave-number space as a function of pressure. 
Finally, Sec. 5 contains concluding remarks. 


\section{Multiple-interface plasma-vacuum model}\label{sec:model}

\begin{figure}[h]
\includegraphics[width=60mm]{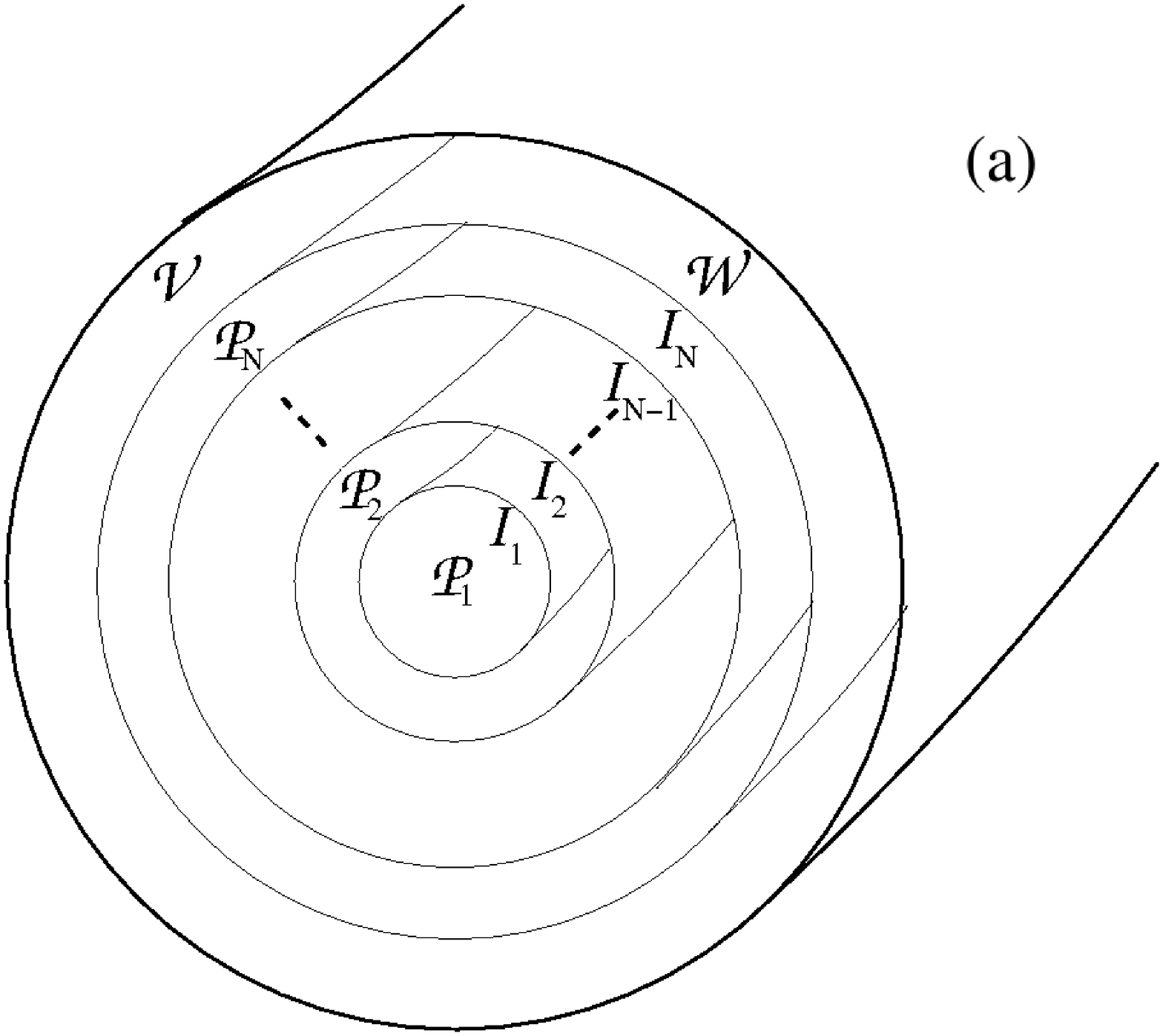} \hspace{1cm}
\includegraphics[width=80mm]{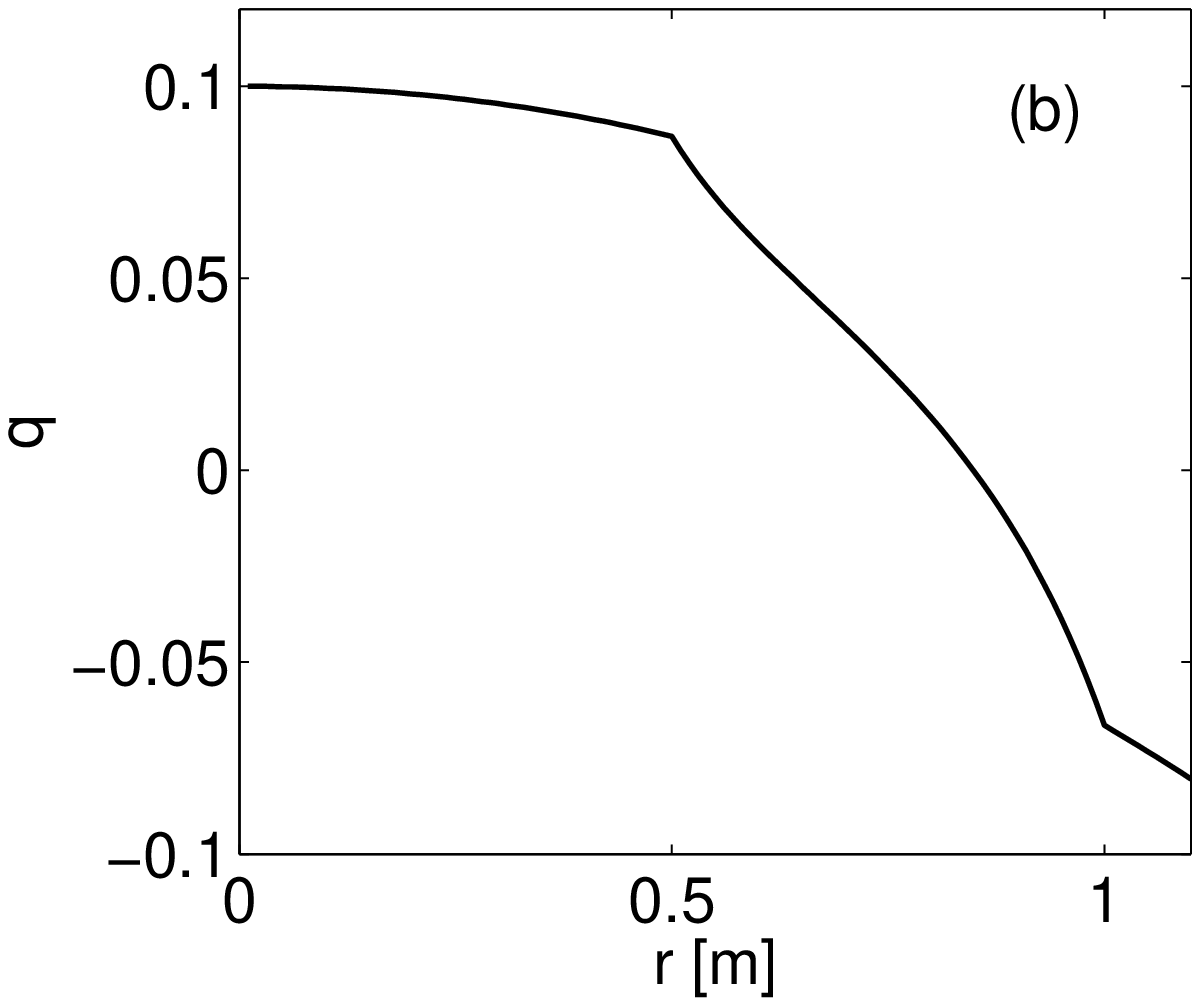}
\caption{\label{fig:geometry} 
\begin{small}
Schematic of magnetic geometry (a), showing ideal MHD barriers $\mathcal{I}_i$, the conducting
wall $\mathcal{W}$, plasma regions $\mathcal{P}_i$ and the vacuum $\mathcal{V}$. Panel (b) shows the $q$ profile used for 
stability studies in Sec. 4, with $\mu_1=2$ (core) and $\mu_2 = 3.6$ (edge).
\end{small} } \end{figure}

The system comprises $N$ Taylor-relaxed plasma regions, each separated by an ideal MHD barrier.  The outermost plasma region is enclosed by a vacuum, 
and encased in a perfectly conducting wall. Figure \ref{fig:geometry}(a) shows the geometry of the system, and introduces 
the nomenclature used to describe the region and interfaces.
The regions $\mathcal{R}_i$ comprise the $N$ plasma regions $\mathcal{R}_1 = \mathcal{P}_1,...,\mathcal{R}_N = \mathcal{P}_N$ 
and the vacuum region  $\mathcal{R}_{N+1} = \mathcal{V}$. 
Each plasma region $\mathcal{P}_i$ is bounded by the inner and outer ideal MHD interfaces $\mathcal{I}_{i-1}$, and $\mathcal{I}_i$ respectively, 
whilst the vacuum is encased by the perfectly conducting wall $\mathcal{W}$.

\subsection{A variational description}\label{subsec:variational}

In previous work \cite{Hole_07} we outlined our variational principle, which lies between that of Kruskal \& Kulsrud \cite{Kruskal_58} 
--- minimization of total energy $W \equiv \int [B^{2}/2 + p/(\gamma - 1)]$ (where $p$ is plasma pressure and $\gamma$ the ratio of specific heats) 
under the uncountable infinity of constraints provided by applying ideal MHD within each fluid element---and the relaxed MHD of 
Woltjer \cite{Woltjer_58} and Taylor \cite{Taylor_74}---minimization of $W$ holding only the two global toroidal and poloidal 
magnetic fluxes, and the single global ideal-MHD helicity invariant $H \equiv \int {\bf A}\cdot{\bf B}$, constant.
In summary, the energy functional could be written
\begin{equation}
W = \sum_{i=1}^N U_i - \sum_{i=1}^N \mu_i H_i/2 - \sum_{i=1}^N \nu_i M_i \label{eq:W_energy}
\end{equation}
where $\mu_i$ and $\nu_i$ are Lagrange multipliers, and
\begin{eqnarray}
\hspace{-2cm} U_i & = & \int_{\mathcal{R}_i} d\tau^3 \left ( { \frac{P_i}{\gamma-1} + \frac{B_i^2}{2 \mu_0} } \right),  \label{eq:U_i}\\
\hspace{-2cm} M_i & = & \int_{\mathcal{R}_i} d\tau^3 P_i^{1/\gamma},  \label{eq:M_i} \\
\hspace{-2cm} H_i & = & \int_{\mathcal{R}_i} d\tau^3\bA \cdot \curl \bA   
                      + \oint_{C_{p,i}^<} \bdl \cdot \bA \oint_{C_{t,i}^<} \bdl \cdot \bA
                      - \oint_{C_{p,i}^>} \bdl \cdot \bA \oint_{C_{t,i}^>} \bdl \cdot \bA . \label{eq:H_i}
\end{eqnarray}
The term $U_i$ is the potential energy, $M_i$ the plasma mass, and $H_i$ the magnetic helicity in each region $\mathcal{R}_i$. 
In Eqs. (\ref{eq:U_i}) - (\ref{eq:H_i}), $d\tau^3$ is a volume element, $\gamma$ the ratio of specific heats, and 
$P_i, B_i$ and $\bA_i$ the equilibrium pressure, magnetic field strength and vector potential respectively.
The superscripts $^>$ and $^<$ denote clockwise and anti-clockwise rotation, respectively.

Setting the first variation to zero yields the following set of equations:
\begin{eqnarray}
\mathcal{P}_i ; & \curl \bB = \mu_i \bB, & \hspace{1cm}  P_i = \mathrm{const.}, \label{eq:Pi} \\
\mathcal{I}_i ; & \bn \cdot \bB = 0,     & \hspace{1cm}  [[ P_i + 1/2 B^2 ]] = 0,  \label{eq:Ii} \\
\mathcal{V}   ; & \curl \bB = 0,         & \hspace{1cm}  \nabla \cdot \bB = 0, \label{eq:V}\\
\mathcal{W}   ; & \bn \cdot \bB = 0,      \label{eq:W}
\end{eqnarray}
where $\bn$ is a unit vector normal to the plasma interface $\mathcal{I}_i$, 
and $[[ x ]] = x_{i+1} - x_i$ denotes the change in quantity $x$ across the interface $\mathcal{I}_i$. 
The boundary conditions, $\bn \cdot \bB =0$, arise because each interface and the conducting wall is assumed to have infinite conductivity.
In turn, these imply the toroidal flux in each plasma region (and the poloidal flux in the vacuum) is constant during relaxation.
Given the vessel with boundary $\mathcal{W}$, the interfaces $\mathcal{I}_i$, and the magnetic field $\bB$,
Eqs. (\ref{eq:Pi})-(\ref{eq:W}) constitute a boundary problem for the plasma pressure $P_i$ in each region $\mathcal{R}_i$.

Minimizing the second variation subject to the constraint of the positive definite normalization $N = \sum_i^N \int_{\mathcal{I}_i} d^2 \sigma |\xi_i|^2 $
yields the following set of equations for the variation in the magnetic field $\bb = \delta \bB$: 
\begin{eqnarray}
\mathcal{P}_i & ; & \curl \bb = \mu_i \bb,    \label{eq:Pb}\\
\mathcal{I}_i & ; & \xi_i^* [[ \bB \cdot \bb ]] + \xi_i^* \xi_i [[ B (\bn \cdot \nabla) B ]] - 
                  \lambda \xi_i^* \xi_i = 0,  \label{eq:Ib1}  \\
              &   & \bn \cdot \bb_{i,i+1} = \bB_{i,i+1} \cdot \nabla \xi_i + \xi_i \bn \cdot \curl (\bn \times \bB_{i,i+1}),   \label{eq:Ib2}\\
\mathcal{V}   & ; & \curl \bb = 0,            \hspace{1cm} \nabla \cdot \bb = 0,\label{eq:Vb} \\
\mathcal{W}   & ; & \bn \cdot \bb  = 0.       \label{eq:Wb}
\end{eqnarray}
Here $\bxi_i$ is the normal displacement of the interface $I_i$, 
and $\lambda$ the Lagrange multiplier of the stability treatment, such that $\lambda<0$ indicates a lower energy state is available. 
Using Eqs. (\ref{eq:Pb})-(\ref{eq:Wb}) the perturbed flux through each region can be found. 
With a suitable Fourier decomposition chosen, Eq. (\ref{eq:Ib2}) solves for the unknown coefficients of the
perturbed field in each region. With substitution,  Eq. (\ref{eq:Ib1}) then becomes a linear eigenvalue equation for $\lambda$. 

\subsection{Tearing mode treatment}\label{subsec:tearing_stability}

A starting point for the treatment of tearing modes is the set of MHD equations: 
\begin{eqnarray}
\frac{\partial \rho}{\partial t} + \vdiv \rho \bv & = & 0, \label{eq:MOTION}\\
\rho \frac{d \bv}{d t}                           & = & \bJ \times \bB - \grad p,    \label{eq:CON}\\
\frac{d}{dt} \frac{p}{\rho^\gamma_{gas}}         & = & 0, \label{eq:ENT}\\
\bE + \bv \times \bB   & = & \eta \bJ, \\
\curl \bE              & = & -\partial \bB/\partial t, \\
\curl \bB              & = & \mu_0 \bJ, \\
\vdiv \bB              & = & 0, 
\end{eqnarray}
being the fluid equation of motion, mass continuity, the adiabatic equation of state, 
Ohm's law, Faraday's law, Ampere's law, and the magnetic mono-pole condition, respectively. The plasma parameters
change across each interface, and across surfaces resonant with perturbations of a given helicity. 

We solve for the plasma parameters for a zero flow plasma (\ie $\bv = 0$) in ``outer'' regions away from the
resonant surfaces where the effects of resistivity are negligible. 
To solve, the field is written $\bB = \grad \chi \times \bh + g \bh$, where $g$ and $\chi$ are scalar functions of position and time 
and $\bh$ is the helical wave-field vector.
Next, $\chi$ and $g$ are expanded as a Fourier perturbation, and solutions to the linearised Beltrami equation found.
The ODE for $\chi_1(r)$, the radial envelope of the linear Fourier perturbation for $\chi$,
integrates to a jump condition in $\chi_1(r)'/\chi_1(r)$ at each interface, expressed in terms of equilibrium parameters. 
The plasma growth rate, obtained by linearising Faraday's law and substituting for $\bE$ as determined by Ohm's law, 
is proportional to $\Delta'= \left [ \chi_1(r)'/\chi_1(r) \right ]^{r_s^+}_{r_s^-} $, 
such that $\Delta'=0$ denotes marginal stability, and $\Delta'>0$ instability. 
The final expression for $\Delta'$ is a function of the equilibrium parameters in the resonant region,
as well as jumps in equilibrium parameters across the interfaces.

\section{MRXMHD cylindrical plasmas}\label{sec:cylinder}

Solutions in an azimuthally-symmetric, axially-periodic cylinder (with axial periodicity length $L=2 \pi R$) are available in Hole \etal \cite{Hole_07}. 
In the cylindrical co-ordinate system $(r, \theta, z)$ they are:
\begin{eqnarray}
\begin{array}{llccc}
\mathcal{P}_1 & : \bB =  \{ 0, & k_1 J_1(\mu_1 r), 			 &  k_1 J_0(\mu_1 r)                    & \}, \label{eq:BP1} \\
\mathcal{P}_i & : \bB =  \{ 0, & k_i J_1(\mu_i r) + d_i Y_1(\mu_i r),     &  k_i J_0(\mu_i r) + d_i Y_0(\mu_i r) & \}, \label{eq:BPi} \\
\mathcal{V}   & : \bB =  \{ 0, & B_{\theta}^V/r,    			 &  B_z^V                               & \}, \label{eq:BV}
\end{array}
\end{eqnarray}
where $k_i, d_i \in \Re $, and $J_0, J_1$ and $Y_0, Y_1$ are Bessel functions of the first kind of order 0, 1, and
second kind of order 0, 1, respectively. The terms $B_{\theta}^V$ and $B_z^V$ are constants.
The constant $d_1$ is zero in the plasma core $\mathcal{P}_1$, because the Bessel functions
$Y_0(\mu_1 r)$ and $Y_1(\mu_1 r)$ have a simple pole at $r=0$. Radius is normalized to the
plasma-vacuum boundary, located at $r=1$. 
The equilibrium is constrained by the $4N+1$ parameters:
\begin{equation}
\{k_1,...,k_N, d_2,...,d_N,  \mu_1, ..., \mu_N, r_1,...r_{N-1}, r_w, B_\theta^V, B_z^V\}, \label{eq:constraints_A}
\end{equation}
where $r_i$ are the radial positions of the $N$ ideal MHD barriers, and $r_w$ is the radial position
of the conducting wall. Equivalent representations, and the mapping between these solutions has been discussed in earlier work [1-3].

\subsection{Stability from a variational principle}\label{subsec:MRXMHD_cylinder}\vspace{6pt}

We have assessed stability using a Fourier decomposition in the poloidal and axial directions for the perturbed field 
$\bb = \curl \ba$ and the displacements $\xi_i$ of each interface. That is, 
\begin{eqnarray}
\bb  =  \F \bb e^{i (m \theta + \kappa z)}, & & \xi_i = X_i e^{i (m \theta + \kappa z)},
\end{eqnarray}
where $m, \kappa$ are the Fourier poloidal mode-number and axial wave-number, and
$\F \bb$ and $X_i$ are complex Fourier amplitudes. Under these substitutions, and after solving for the field in each region, 
Eq. (\ref{eq:Ib1}) reduces to an eigenvalue matrix equation $ \bldeta \cdot \bX = \lambda \bX $
with column eigenvector $\bX = (\xi_1, ..., \xi_{N})^T$, eigenvalue $\lambda$, and $\bldeta$ a $N \times N$ tridiagonal real matrix. 

\subsection{Tearing mode stability}\label{subsec:tearing}

In the helical coordinate $u=m \theta + \kappa z$, a divergence-less $\bB$ can be written
\begin{equation}
\bB(r, u) = \grad \chi(r,u) \times \bh + g(r, u) \bh, \label{eq:Btearing}
\end{equation}
where $\chi$ is a the helical flux, and $g$ a helical field. The vector $\bh$ is defined by 
$\bh = f(r) \grad r \times \grad u$, where $f(r)=r/(m^2 + k^2 r^2)$ is a metric term.
As in Tassi \etal we search for helical perturbations of the form 
\begin{eqnarray}
\chi(r,u,t) = \chi_0(r) + \chi_1(r) e^{\gamma t + i u}, &  & g(r,u,t) = g_0(r)    + g_1(r) e^{\gamma t + i u}. 
\end{eqnarray}
In this representation, resonant surfaces are those for which $\chi_0'(r) = 0$.  
The equilibrium field satisfies the Beltrami equation, giving rise to $\mu = g_0'(r)/\chi_0'(r)$, 
such that the rotational transform can be written
\begin{equation}
\iotabar = -\frac{R}{r} \times \frac{r \kappa g_0(r)/\chi_0'(r) + m}{m g_0(r)/\chi_0'(r) - r \kappa}. \label{eq:iotabar}
\end{equation}
By writing the incompressible velocity field in a similar form to Eq. (\ref{eq:Btearing}), and expanding continuity to first order, 
it is possible to show perturbations in the flow, pressure and mass density do not affect marginal stability.

In each of the plasma regions, projections of the linearised Beltrami equation along $\bh$ and $\grad r$ yield
\begin{eqnarray}
\hspace{-2cm} g_1 = g_0'(r)/\chi_0'(r) \chi_1(r),  && \label{eq:g1}\\
\hspace{-2cm} \chi_0(r) \left [ {
\chi_1''(r) + 
\frac{f'(r)}{f(r)} \chi_1'(r) + 
\left ( { \mu^2 - \frac{1}{r f(r)} + \frac{g_0(r)}{\chi_0'(r)} \mu' + \frac{2 m \kappa}{m^2 + \kappa^2 r^2} \mu } \right ) \chi_1(r) } \right ] = 0,  && \label{eq:chi1}
\end{eqnarray}
where $\mu'$ vanishes everywhere except at $r=r_{step}$, where it becomes singular. These are identical to Eqs. (27) and (29) of
Tassi \etal. Equation (\ref{eq:chi1}) reduces to a Bessel or modified Bessel equation in the transformed variables 
$x_i= \sqrt{|\mu_i^2 - \kappa^2|} r$ and $x_V = |\kappa| r$. In the $i$'th region, and either side of the resonant surface, 
solutions are different combinations of Bessel or modified Bessel functions with undetermined coefficients $\zeta_i, \Lambda_i$, 
$\zeta_{i-s}, \Lambda_{i-s}$ and $\zeta_{i+s}, \Lambda_{i+s}$, respectively. 
As only the ratio $\chi_1'(r)/\chi_1(r)$ appears in  $\Delta'$, its value is unaffected by normalizing $\zeta = 1$ in each interval and region.
The requirement of boundedness at $r=0$, and the presence of perfectly conducting wall implies
\begin{equation}
\chi_1(0) = \chi_0, \hspace{1cm} \chi_1(r_w) = 0. \label{eq:chi1BC}
\end{equation}
Noting that the perturbed flux must be continuous across each interface, Eq. (\ref{eq:chi1}) can then be integrated about each interface to yield
\begin{eqnarray}
\left [ \left [  {{ \chi_0'(r) \frac{\chi_1'(r)}{\chi_1(r)} }}  \right ] \right ]  = 
\left [ \left [ {{ -\chi_0'(r) \frac{g_0(r)}{\chi_0'(r)} \mu }} \right ] \right ]\label{eq:chi1_red}
\end{eqnarray}
The parametric dependence can also be examined by solving for $\chi_0(r)$ and using Eq. (\ref{eq:iotabar}) to eliminate $\frac{g_0(r)}{\chi_0'(r)}$.
Solving equilibrium for $\chi_0(r)$ gives
\begin{eqnarray}
\chi_0'(r) = B \sqrt{\frac{ m^2 + \kappa^2 r^2}{\left ( {\frac{-m R/r + \kappa R \iotabar r/R}{\iotabar m + \kappa R} } \right )^2 + 1}} =  F(B, r, R, m, \kappa). 
\label{eq:Fdef}
\end{eqnarray}
Finally, eliminating $\chi_0(r)$ and $\frac{g_0(r)}{\chi_0'(r)}$, Eq. (\ref{eq:chi1_red}) can then be rewritten
\begin{equation}
\left [ \left [ { F(B, r, R, m, \kappa) x \frac{\chi_1'(x)}{\chi_1(x)}}  \right ] \right ]  = 
\left [ \left [ { G(B, \mu, \iotabar, r, R, m, \kappa) }  \right ] \right ], \label{eq:dlogchi_1}
\end{equation}
where 
\begin{equation}
\hspace{-1cm}
G(B, \mu, \iotabar, r, R, m, \kappa) = 
r \mu F(B, r, R, m, \kappa) \times \frac{m R/r - \kappa R \iotabar r/R}{\iotabar m + \kappa R}. \label{eq:Gi}
\end{equation}
With $\zeta=1$ everywhere, the tearing mode parameter $\Delta'$ is a function of $\Lambda$ in each interval, 
which is uniquely determined by the set of constraints given by Eq. (\ref{eq:dlogchi_1}) at each interface. 
That is, the inner boundary condition (\ref{eq:chi1}) yields $\Lambda_1 = 0$.  If for instance the resonant surface lies in $\mathcal{R}_2$, 
Eq. (\ref{eq:dlogchi_1}) evaluated at interface $\mathcal{I}_1$ and  $\mathcal{I}_2$ solves for $\Lambda_{2-s}$ in terms of $\Lambda_1$ 
and $\Lambda_{2+s}$ in terms of $\Lambda_V$, respectively. The conducting wall boundary condition solves for $\Lambda_V$.  

Changes in the field strength $B$ at any interface enter Eq. (\ref{eq:chi1_red}) through the solution to $\chi_0('r)$, given by Eq. (\ref{eq:Fdef}). 
Stability is hence a property of the rotational transform, the position of the barriers, the Lagrange multipliers, the rotational transform, and any jumps
in the pressure or rotational transform across the interfaces. 
Our working reduces to Tassi \etal in the limit of no pressure, field or rotational transform jumps across the interfaces, and no vacuum.

\section{Stability for an RFP-like configuration}\label{sec:pressureless}

We have compared stability conclusions using variational and tearing mode treatments for an illustrative configuration. 
The example chosen is guided by earlier detailed working \cite{Hole_07}, where the core Lagrange multiplier was $\mu_1 = 2$. 
The first interface is placed at $r=0.5$, and the axial periodicity length chosen to be $20 \pi$, such that 
the effective aspect ratio $r/R=1/10$ is small. The jump in safety factor between the internal interface and the plasma-vacuum boundary
has been chosen to resemble Hole \etal, subject to the different $R$ values used for the two treatments ($R=1/(2 \pi)$ in Hole \etal).  
We have used $\mu_2 = 3.6$, which requires $d_2/k_2 = 0.77$ for the rotational transform profile to be continuous. 
A second motivation for this choice is the similarity to $q$ profiles of high confinement reverse field pinches, 
such as the Madison Symmetric Torus \cite{Chapman_02} and RFX-mod \cite{Ortolani_06}, although the change in $\mu$ is greatly exaggerated in this work.
The plasma pressure is selected by the parametrization $\beta_1 = p_1/(B_V^2/2 \mu_0), \beta_2 = p_2/(B_V^2/2 \mu_0)$. 
Except for the final scan over $\beta$, a pressure-less plasma is assumed (\ie $\beta_1 = \beta_2 = 0$).
Figure \ref{fig:geometry}(b) shows the $q$ profile for the chosen equilibrium, where $q = 1/\iotabar$.
   
\begin{figure}[h]
\includegraphics[width=80mm]{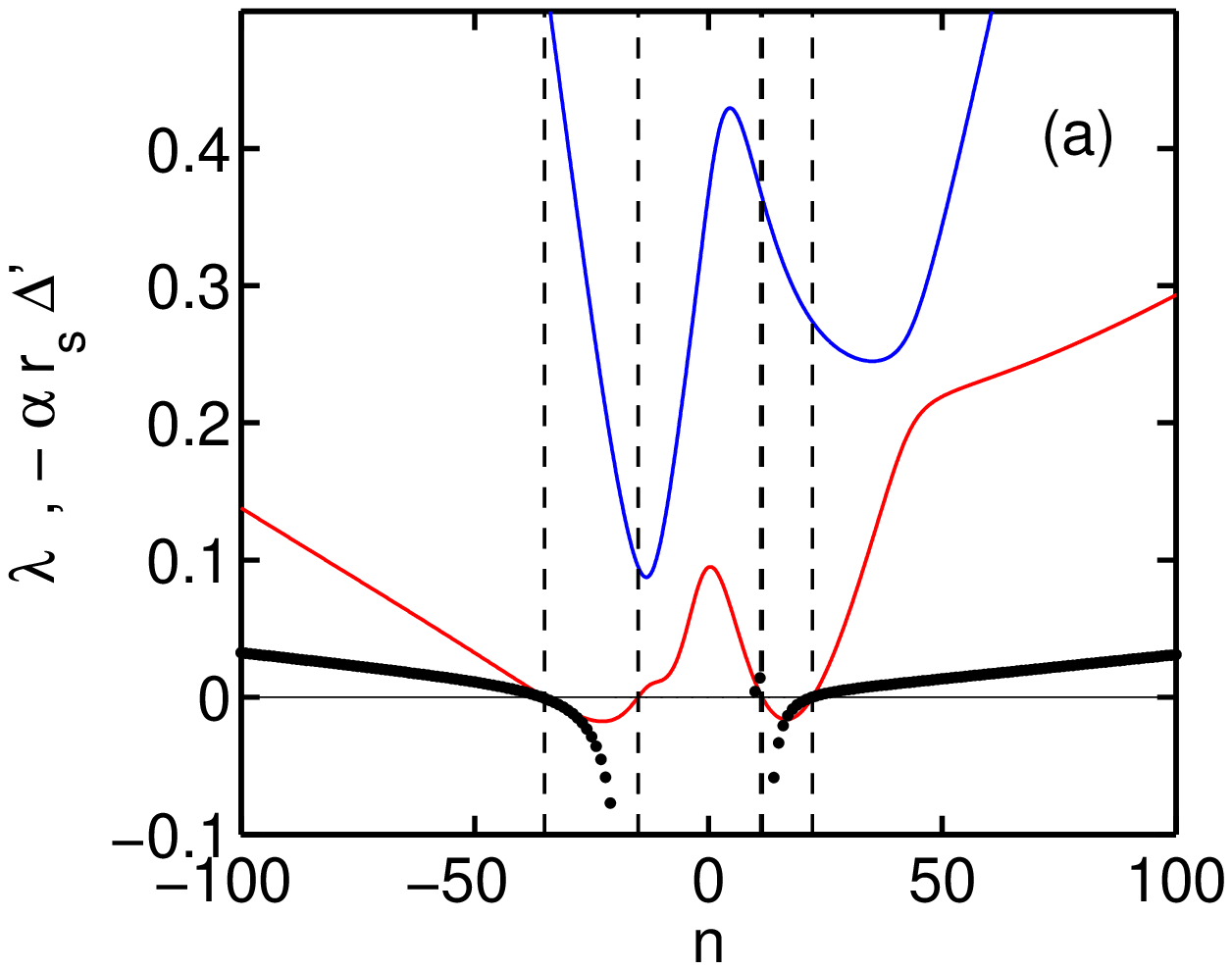}
\includegraphics[width=80mm]{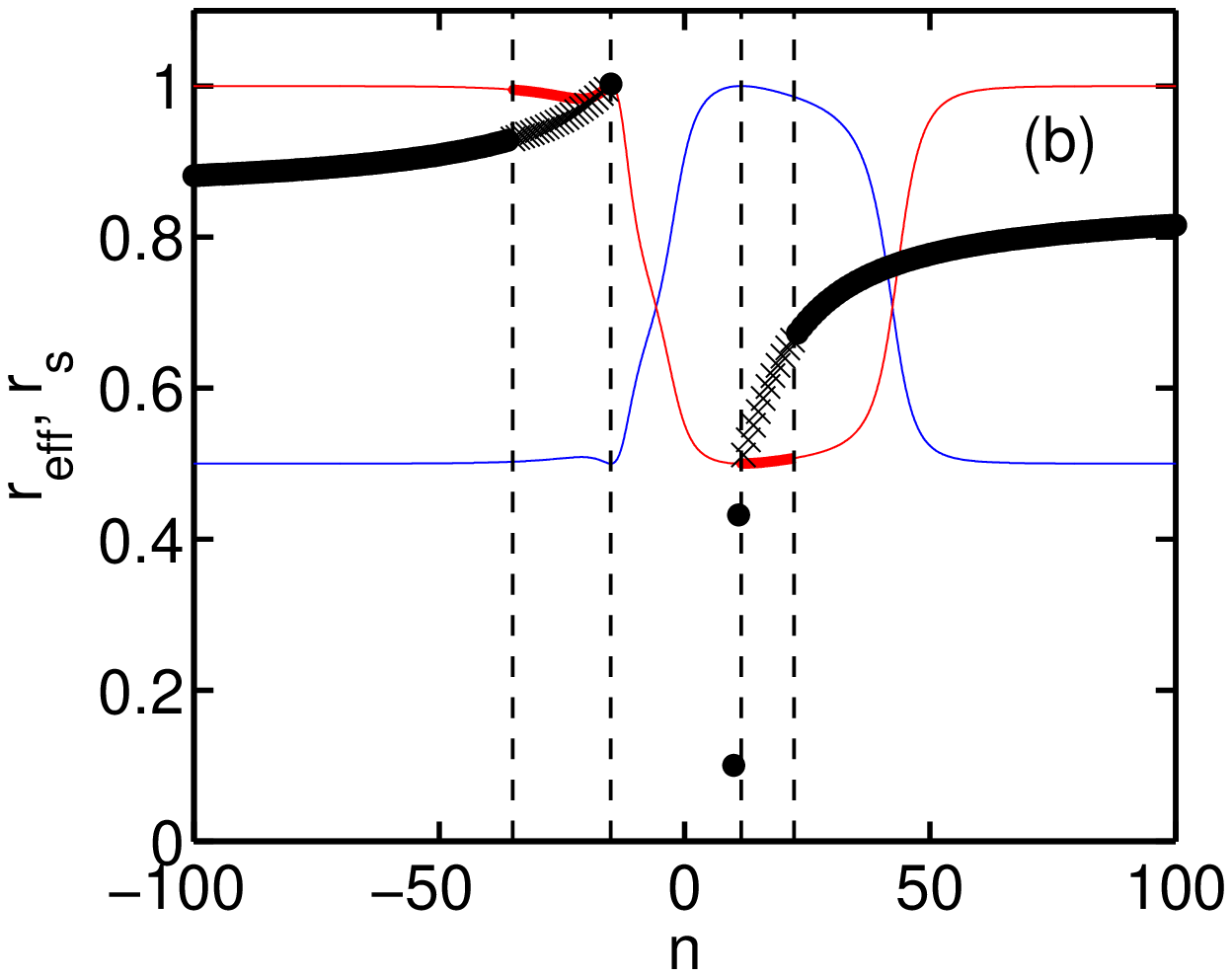}
\caption{\label{fig:disp_curve} 
\begin{small}  
Dispersion curves (a) and mode localization (b) of $m=1$ modes of a pressure-less MRXMHD plasma with $q$ profile given by 
Fig. \ref{fig:geometry}(b). In panel (a), the solid lines are eigenvalues ($\lambda$) of the MRXMHD treatment, and represent different eigenfunctions,
while the points are values of $-0.002 \times r_s \Delta'$ determined from tearing mode analysis of Sec. \ref{subsec:tearing}.
The vertical dashed lines correspond to zeros in $\lambda$. 
Panel (b) shows the resonant surfaces $r_s$ of tearing modes (points), and effective localization $r_{s, eff}$ of modes using the variational treatment (solid line).
The solid points and cross-hairs denote stable and unstable tearing modes, respectively. 
The heavy solid line denotes solutions for which $\lambda<0$, and the dashed vertical lines correspond to marginal stability, $\lambda=0$.
\end{small} }\end{figure}

Figure \ref{fig:disp_curve}(a) is a dispersion curve for $m=1$ modes, showing $\lambda$ computed  using the variational treatment, 
and $-r_s \Delta'$ computed for modes resonant within the plasma. 
Marginal stability corresponds to $\lambda = 0$ and $\Delta'=0$: these overlap identically in Fig.\ref{fig:disp_curve}(a).
Modes with $n=-16$ and $n=12$ corresponds to a perturbation near-resonant with the 
outer and inner interfaces ($q(r_2) \approx -1/16$ and $q(r_1) \approx 1/12$) respectively.  

In our variational treatment, we have prescribed no relationship between $\xi$ and $\bb$ in the relaxed regions.
As such, excepting at the ideal interfaces, field line resonance in such plasmas is not explicitly resolved. 
Expressions can however be constructed which provide an estimate of the  localization of the mode 
$r_{s, \mathrm{eff}}$, and a convenient choice is $ r_{s,\mathrm{eff}}^2 = \sum_{i=1}^N (r_i X_i)^2$,
where the eigenvectors are normalised such that $\bX \cdot \bX^H=1$, with $H$ the Hermitian.

\begin{figure}[h]
\includegraphics[width=80mm]{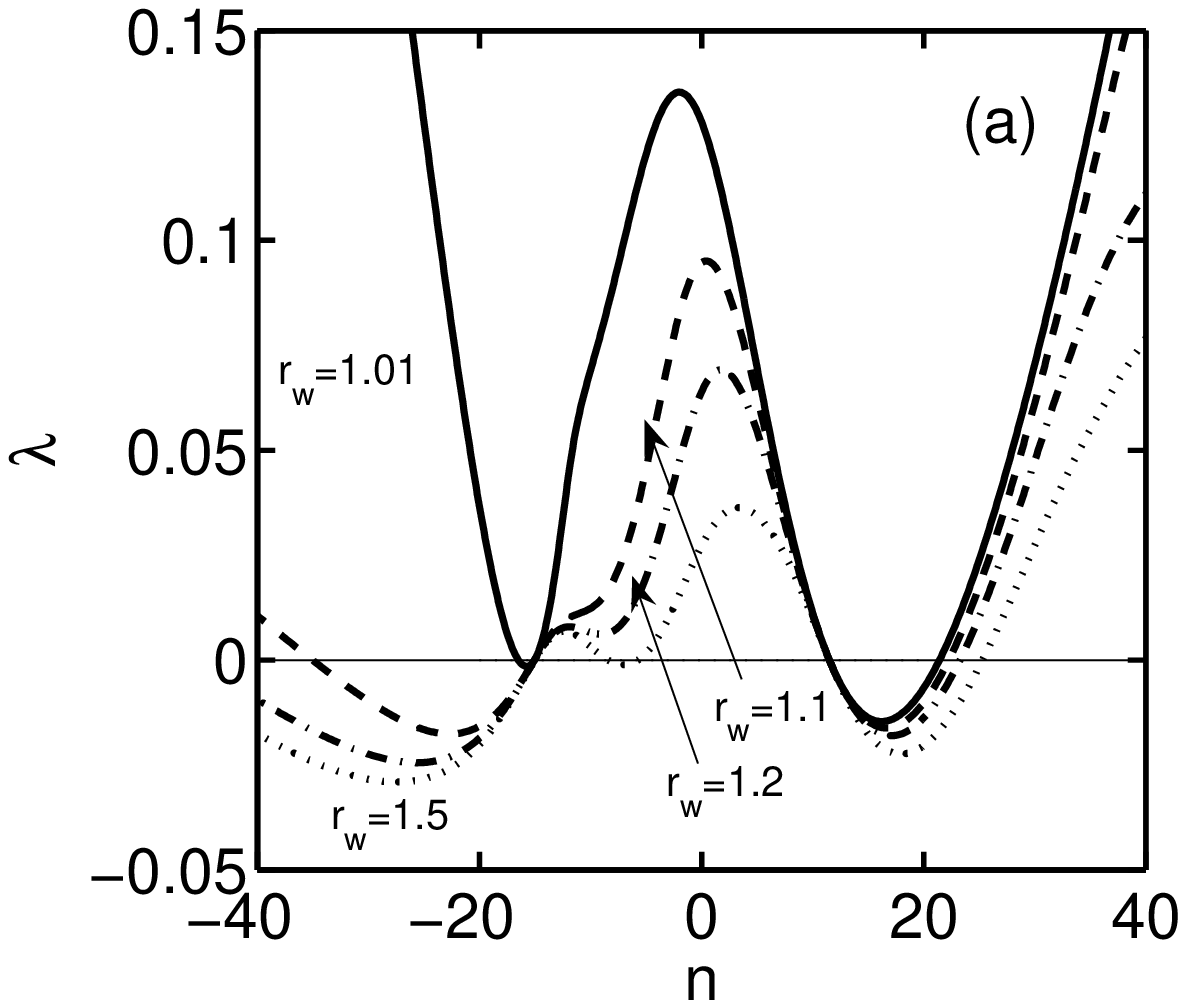}
\includegraphics[width=80mm]{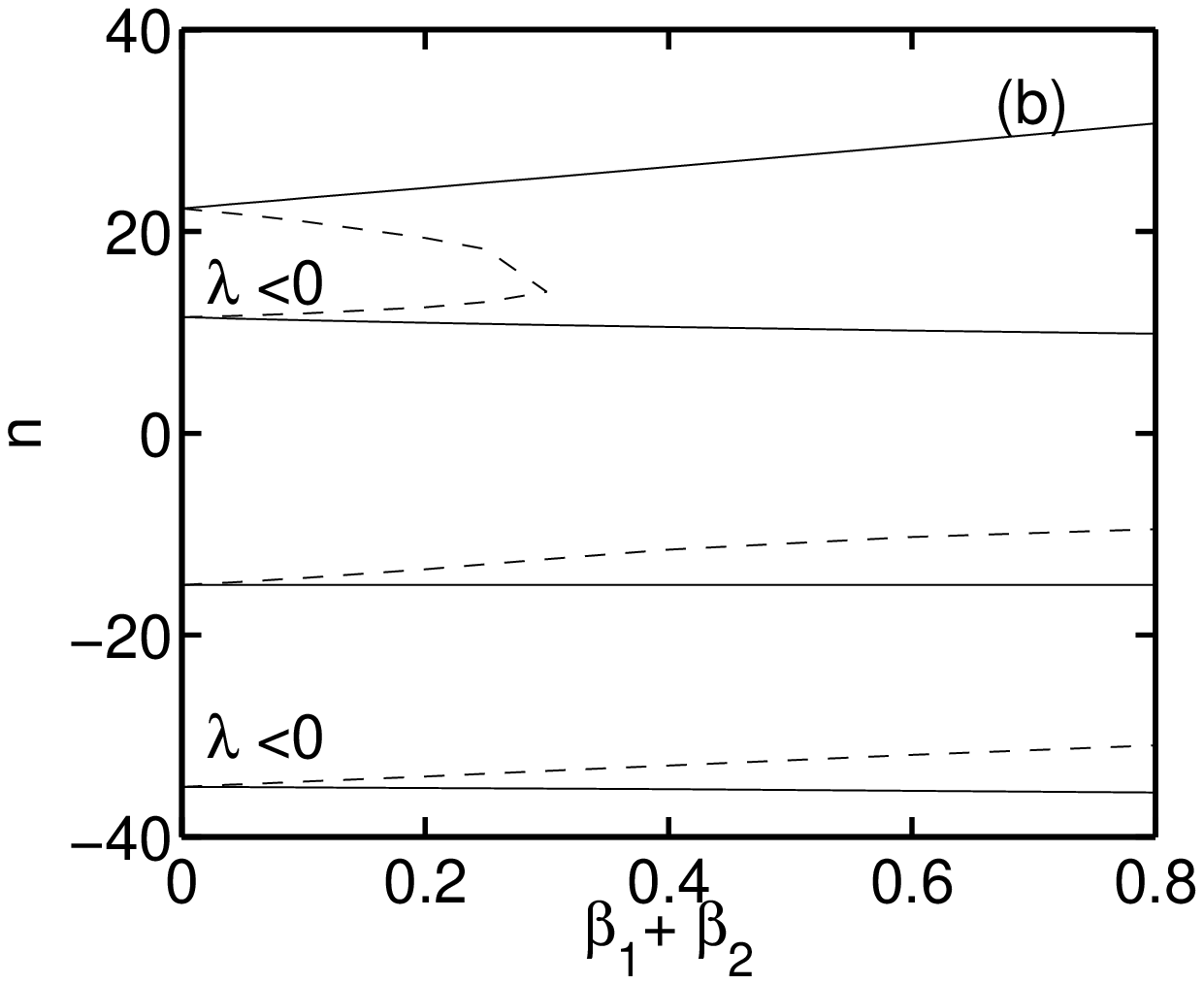}
\caption{\label{fig:wall} \begin{small} Wall stabilization (a) and marginal stability pressure dependence (b) of MRXMHD plasmas unstable to $m=1$ modes.
Panel (a) shows the dispersion curves of MRXMHD plasmas unstable to $m=1$ modes as a function of $n$ for different conducting wall radii. 
In panel (b) marginal stability $n-\beta$ space is shown for different pressure profile configurations.
The solid lines are for the pressure in the outer region set to zero ($\beta_2=0$), while the dashed line corresponds to zero core pressure ($\beta_1 = 0$).
\end{small}} 
\end{figure}

Figure \ref{fig:disp_curve}(b) shows a comparison of $r_{s, \mathrm{eff}}$ to $r_s$, in which modes unstable to variational and tearing modes have been identified.
Agreement between $r_{s,\mathrm{eff}}$ and $r_s$ is qualitatively good in the interval over which the plasma in unstable, and excellent near the interfaces.
The $n=-16$ and $n=9$ modes are near resonant with the outer and inner interface, respectively.  
As shown in Fig. \ref{fig:wall}(a), a stability scan with wall radius indicates that in the limit $r_w \rightarrow 1$, modes for $n<0$ are wall-stabilized.
In the limit that the outer interface is made resonant with the $n=-16$ tearing mode
(for example by changing $R$), $\Delta' \rightarrow \infty$. This mode is the current driven external kink of ideal MHD. 
Conversely, the unstable range for $n \ge 12$ is only very weakly affected by the wall position. If the inner 
interface is mode resonant with the perturbation, $\Delta' \rightarrow \infty$, and the mode is ideal unstable. 
This is the internal kink of ideal MHD.

Recently, Mills \etal \cite{Mills_08} demonstrated that one can unify ideal and relaxed variational treatments by extending the relationship between $\bb$ and $\bxi$ 
through the Newcomb gauge $\ba = \bxi \times \bB$. If this variational treatment is followed, rational surfaces do explicitly enter the expression for 
$\xi$ as derived from $b_r$, the variation in the radial part of the magnetic field, and so both relaxed and tearing modes become localised at the resonant surface $r_s$.

The findings of Fig. \ref{fig:disp_curve}, obtained for a pressure-less plasma, agree with that of Furth \etal \cite{Furth_73}, who showed that for cylindrical pressure-less 
plasma with no vacuum, $\Delta'$ is linear with the second variation in the magnetic energy driving the tearing mode.  
We have also compared stability conclusions drawn from variational and tearing mode treatments as a function of $\beta$.
We find that while both variational and tearing mode treatments have the same stability trends with $\beta$, 
the marginal stability limit of the variational treatment (for a given $m$ and $n$) is lower than that of tearing modes.
A study is ongoing into the cause of this discrepancy, as well as formally relating $\delta^2 W$ to $\delta'$ 

Finally, Fig. \ref{fig:wall}(b) is a plot of the marginal stability boundary ($\lambda=0$) in $n-\beta$ space for $m=1$ eigenmodes of the variational treatment. 
The two pressure profile configurations that have been studied are $\beta_1=0$ and $\beta_2 = 0$. 
Trends in the marginal stability boundary can be understood by relating the radial location of the mode resonant surface to the
analog of radial pressure gradient in the MRXMHD model: the sign and magnitude of nearby pressure jumps. 
For $n>0$ modes resonant near the first interface, an increasing core pressure increases the pressure drop across the first interface,
and so destabilises the plasma.  
Conversely, increasing the edge pressure leads to a pressure jump across the first interface, and so stabilises the internal modes. 
For $\beta_2>0.3$ all $m=1$ internal modes ($n>0$) are completely stabilised. 
For the $n<0$ modes resonant near the edge, changes in the core pressure have little effect, while
increasing the edge pressure destabilises the plasma.  

\section{Conclusions}\label{sec:conclusions}

We have computed the stability of multiple relaxation region MHD (MRXMHD) plasmas using both a variational and a tearing mode treatment,
evaluated in a periodic cylindrical configuration. 
The marginal stability conclusions of the two treatments for a zero $\beta$ plasma, as well as the trends with $\beta$, appear to 
be identical, in agreement with earlier analytic working by Furth \etal \cite{Furth_73}.
Some discrepancy exists between the marginal stability boundaries of variational and tearing mode treatments for nonzero $\beta$, 
with the stability limit of variational plasmas lower than that of tearing modes. A study is underway to resolve this discrepancy.

The overarching aim of this work is to elucidate the nature of perturbations available to MRXMHD equilibria, which in turn, are
motivated by our quest for mathematically rigorous solutions of ideal MHD force balance in 3D geometry. 
Our working builds of Tassi \etal to nonzero $\beta$ multiple region relaxed plasmas
with vacuum, and complements Mills \etal, who demonstrated that 
that one can unify ideal and relaxed variational treatments through the Newcomb gauge. 
Combined, these results suggest that MRXMHD configurations which are stable to ideal perturbations plus tearing modes are automatically in a 
stable equilibrium state. 

In ongoing work we are  developing a faster numerical algorithm for the construction of 3D MRXMHD plasmas, and exploring the
maximum pressure jump an interface can support before it is destroyed by chaos.\\


\noindent {\bf Acknowledgments}
The authors would like to acknowledge the support of the Australian Research Council, through grant DP0452728. 

\section*{References}
 
\bibliographystyle{unsrt}
\bibliography{3D_MHD}

\end{document}